\documentclass[12pt,a4paper]{article}
\usepackage[T1]{fontenc}
\usepackage{graphicx}
\usepackage{graphics}
\usepackage{feynmp}
\usepackage{hyperref}
\usepackage{amsmath}
\usepackage{ifthen}

\newcommand{\beq}[1]{\begin{equation}{\label{#1}}}
\newcommand{\eeq}[0]{\end{equation}}
\newcommand{\barr}[0]{\begin{array}}
\newcommand{\earr}[0]{\end{array}}
\newcommand{\eq}[1]{eq.(\ref{#1})}
\newcommand{\Eq}[1]{Equation(\ref{#1})}

\renewcommand{\beq}[1]{\begin{equation}{\label{eq:#1}}\begin{aligned}\hspace*{0pt}}
\renewcommand{\eeq}{\end{aligned}\end{equation}}
\renewcommand{\eq}[1]{eq.\,(\ref{eq:#1})}
\renewcommand{\Eq}[1]{Equation\,(\ref{eq:#1})}
\usepackage{graphicx}
\newcommand{\Text}[1]{\quad\mbox{#1}\quad}
\begin{document}
\begin{flushright}
\vspace*{-10mm}
September 6, 2023\\
\end{flushright}
\begin{center}
{\LARGE\bf Regularised unfolding with a\\[2mm]
           discrete-valued penalty function
           \footnote{published open access article at
                     \href{https://doi.org/10.1016/j.nima.2023.168566}
                          {https://doi.org/10.1016/j.nima.2023.168566}}}\\[8mm]
Michael Schmelling\footnote{email: {\tt michael.schmelling@mpi-hd.mpg.de}}\\[1mm]
{\it Max Planck Institute for Nuclear Physics, Heidelberg, Germany}
\end{center}

\begin{abstract}
Regularisation allows one to handle ill-posed inverse problems. Here 
we focus on discrete unfolding problems. The properties of the results 
are characterised by the consistency between measurements and unfolding 
result and by the posterior response matrix. We introduce a novel 
regularisation scheme based on a discrete-valued penalty function 
and compare its performance to that of a simple cutoff-regularisation. 
The discrete-valued penalty function does not require a regularisation 
parameter that needs to be adjusted on a case-by-case basis. In toy 
studies very satisfactory results are obtained.
\end{abstract}

\section{Introduction}
A common problem in the analysis of experimental data is that an actual 
measurement is not only subject to statistical fluctuations, but in 
general differs from the underlying true value (or values), from now on 
referred to as ``truth'', also due to systematic shifts between true and 
measured quantity, efficiency losses and finite resolution of the detector 
system. Considering the one-dimensional case and assuming that the mapping 
between the truth and the asymptotically expected measurement is linear, 
it is described by the Fredholm integral equation of first kind,
\beq{fred1}
       g(x) = \int\,dy\,R(x,y)\,f(y)  \;,
\eeq
where the response function $R(x,y)$ relates the true density $f(y)$ to the
measurement $g(x)$. The integral is over the support of $f(y)$. 
The function $R(x,y)$ parametrises the effects mentioned above and is 
assumed to be independent of $f(y)$. 

In the following we will address the discrete problem, where the response 
function is replaced by a response matrix and the densities $g(x)$ and $f(y)$ 
by discrete distributions $a_i$, $i=1,\ldots,n_a$ and $b_j$, $j=1,\ldots,n_b$, 
respectively. In order to obtain a discrete unfolding problem we define
\beq{disc}
    b_j = \int_{y_{j-1}}^{y_j} dy\,f(y)
    \Text{,}
    R_{ij} = \frac{1}{y_{j}-y_{j-1}}\int_{x_{i-1}}^{x_i} dx \int_{y_{j-1}}^{y_j} dy\; R(x,y)
    \Text{and}
    a_i = \sum_{j=1}^{n_b} R_{ij} b_j \;.
\eeq
If the binning is sufficiently fine, such that the curvature of $f(y)$ or 
$R(x,y)$ over a bin $\Delta y$ can be ignored, one has
\beq{}
    a_i \approx \int_{x_{i-1}}^{x_i} dx\;g(x) 
\eeq
and \eq{disc} becomes a discrete approximation of \eq{fred1} with the components 
of $a$ and $b$ representing the bin-integrated densities $g(x)$ and $f(y)$. 
In the following individual components are addressed by an index. When the index 
is omitted the entire object is referred to.

The number of bins $n_a$ and $n_b$ representing the observed and the true 
distribution can in general be different. The quantities that are used to infer 
an estimate $\hat{b}$ of the true distribution are the measurements $\hat{a}$, 
their covariance matrix $C$ and the response matrix $R$. Here $\hat{a}$ is 
assumed to be an unbiased estimate of the expectation values $a$, with in 
general $\hat{a}\neq a$, i.e. repeating an experiment will result
in different measurements $\hat{a}$. In contrast, $C$ and $R$ are fixed
and assumed to be known without uncertainty.

The discrete unfolding problem \eq{disc} is model-independent in the sense that 
the true distribution is constructed from a complete basis consisting of $n_b$ 
independent basis vectors to parametrise the bin contents. These can be the $n_b$ 
bins used to represent the distribution or linear combinations thereof. The 
alternative would be to have a parametric model of the truth with the number 
of parameters $n_p$ satisfying $n_p\ll n_a$ \cite{James2021}. Those parameters 
then can be estimated by adjusting them such that a forward-folded model 
provides a best fit of the data.

\section{Regularisation and posterior response}
Since the unfolding problem is linear and considering the case $n_a\geq n_b$, 
where the measurements constrain or over-constrain the true distribution, an 
unbiased estimate with minimal variance \cite{Cowan1998} is obtained by 
minimising 
\beq{chi2}
        \chi^2(b,\hat{a}) = (\hat{a} - R b)^T C^{-1} (\hat{a} - R b) \;.
\eeq
The estimate $\hat{b}$ and its covariance matrix $C(\hat{b})$ are given by
\beq{lsqfit}
    \hat{b} = (R^T C^{-1} R)^{-1} R^T C^{-1} \hat{a} 
    \Text{and}
    C(\hat{b}) = (R^T C^{-1} R)^{-1}
    \Text{with}
    \chi^2_{\min} = \chi^2(\hat{b}) \;,
\eeq 
where asymptotically $\chi^2_{\min}$ follows a $\chi^2$-distribution with 
$n_a-n_b$ degrees of freedom and $\chi^2_{\min}$ provides a quantitative 
measure for the consistency between data and unfolded distribution 
\cite{Cowan1998}.  

In typical applications $R^T C^{-1} R$ is an ill-conditioned matrix, with 
the consequence that $\hat{b}$ is dominated by statistical fluctuations. 
Tikhonov regularisation \cite{Tikhonov1943} does address this problem 
by solving a modified minimisation problem with the cost function 
\beq{F}
      F(b) = \chi^2(b,\hat{a}) + w\,S(b) \;,
\eeq 
where a smoothing function $S(b)$ penalises unwanted solutions. Note 
that the $\chi^2$-function depends both on the solution $b$ and the 
measurements $\hat{a}$, whereas $S(b)$ is a function of only $b$.
The regularisation parameter $w$ allows one to adjust the regularisation 
strength. For $w=0$ the unbiased and usually unstable solution is
recovered, for $w\to\infty$ the data are ignored and the solution 
$\hat{b}$ that minimises $S(b)$ is the one that also minimises $F(b)$. 
An added benefit is that in the regularised approach also 
under-constrained problems $n_a<n_b$ have a well defined solution,
since the regularisation term $w S(b)$ lifts the degeneracy of the 
$\chi^2$ function, which for an $n_b-n_a$ dimensional subspace in 
$b$ satisfies $\chi^2=0$.

The key to the interpretation of the unfolding results is the sensitivity 
of $\hat{b}$ to changes in the measurements $\hat{a}$, in the following
described by the matrix $M$ with matrix elements
\beq{Mkl}
     M_{kl} = \frac{d\hat{b}_k}{d\hat{a}_l} \;.
\eeq
In error propagation \cite{Cowan1998} $M$ determines the covariance 
matrix $C(\hat{b})$ of the unfolding result via
\beq{Chat}
    C(\hat{b}) = M C M^{T} \;.
\eeq
If the second derivative of $S(b)$ is a constant matrix, the solution $\hat{b}$ 
is a linear function of $\hat{a}$ and the transformation of the covariance matrix
\eq{Chat} is exact. In the original Thikhonov regularisation scheme this is the
case. For a general regularisation function that is not a quadratic form in $b$, 
\eq{Mkl} is a linear approximation. For sufficiently strong regularisation this 
approximation will be quite good, as can be tested by comparing an error estimate 
based on \eq{Chat} to a bootstrap estimate which also accounts for higher order 
terms. Given $M[n_b,n_a]$, where the square brackets indicate the dimensions, 
two square matrices can be constructed by multiplication with the response 
matrix $R[n_a,n_b]$:
\beq{PQ}
     P = M\,R  
     \Text{and}
     Q = R\,M \;.
\eeq

For the interpretation of $P$ we assume that the measurements $\hat{a}$ track 
the expectation values $a$, i.e.~while in general one has $\hat{a}\neq a$, we 
assume that if $a$ were shifted by $da$ then the actual measurements would 
have come out also shifted with $d\hat{a}=da$. From $a= R\,b$ then follows
$R_{il}=da_i/db_l=d\hat{a}_i/db_l$  and thus
\beq{}
    P_{kl} = \sum_{i=1}^{n_a} M_{ki} R_{il} 
           = \sum_{i=1}^{n_a} \frac{d\hat{b}_k}{d\hat{a}_i} 
                              \frac{d\hat{a}_i}{db_l} 
           = \frac{d\hat{b}_k}{db_l}  \;.
\eeq
The matrix $P$ describes how the unfolding result $\hat{b}$ changes under a change
of the true distribution $b$. If it is not a unit matrix, then the unfolding 
result $\hat{b}$ is not an unbiased estimator of the truth. In analogy to the 
response matrix $R$, whose matrix elements $R_{il}$ describe the expected change 
in a measurement $a_i$ under a change in $b_l$, the matrix element $P_{kl}$ 
describes by how much bin $\hat{b}_k$ of the unfolded distribution changes when 
the truth $b_l$ is varied. The matrix $P$ thus behaves like a response matrix, 
which describes how the unfolding result is distorted with respect to the truth, 
and we will refer to it as ``posterior response matrix''.  It provides 
a quantitative description of the fact that unfolding methods in most 
practical applications cannot fully undo smearing effects and at best 
only achieve an improvement of the resolution function \cite{Zhigunov1983}.

Comparing \eq{Chat} and \eq{PQ}, one sees that properties of the posterior 
response matrix are also visible in the covariance matrix of the bins of 
the unfolded distribution and vice versa. If the same binning is used for 
the true and the observed distribution, then the trivial case $M=1$, 
i.e.~no unfolding, leaves the detector response and covariance matrix 
unchanged, while perfect unfolding $M=R^{-1}$ leads to $P=1$ and a 
solution with usually huge uncertainties and almost full anti-correlation 
between adjacent bins. An example is given e.g. in \cite{Cowan1998}. 
The anti-correlations are a direct consequence of the fact, that the inverse 
of a typical response matrix with only positive matrix elements has 
alternating signs between adjacent matrix elements.

The interpretation of $Q$ follows from considering what happens 
when the unfolding result $\hat{b}$ is multiplied by $R$, i.e.~when 
checking how well $\hat{b}$ reproduces the data $\hat{a}$. The product 
$R\hat{b}$ is a smoothed estimate $\hat{\hat{a}}$ of $\hat{a}$, 
and one has $R_{ik} = d\hat{\hat{a}}_i/d\hat{b}_k$. It follows
\beq{}
    Q_{kl} = \sum_{i=1}^{n_b} R_{ki} M_{il} 
           = \sum_{i=1}^{n_b} \frac{d\hat{\hat{a}}_k}{d\hat{b}_i} 
                              \frac{d\hat{b}_i}{d\hat{a}_l} 
           = \frac{d\hat{\hat{a}}_k}{d\hat{a}_l}  \;.
\eeq
The matrix $Q$ describes how tightly the fluctuations in $\hat{a}$ couple 
to the values $\hat{\hat{a}}$ that result from the estimate $\hat{b}$ of 
the true distribution. Thus it is a measure of the regularisation strength 
and we will refer to it as ``regularisation matrix''. Only if $Q$ is a unit 
matrix, then the measurements directly determine the unfolding result. In 
general some damping will occur.
     
An invariant quantity that characterises the result of the unfolding
procedure is the trace $T$ of the matrices $P$ and $Q$
\beq{}
       {\rm Tr}(M R) = {\rm Tr}(R M) = T  \;.
\eeq
Because the trace is invariant under cyclic permutations, the traces of 
$P$ and $Q$ are the same. With respect to $Q$ the trace $T$ is a measure 
for how many of the bins $\hat{a}$ effectively contribute to the result 
$\hat{b}$. This implies that heuristically the number of degrees of freedom 
for the $\chi^2$-measure that quantifies the agreement between $\hat{a}$ 
and $\hat{b}$ is $N_{df}=n_a-T$, and an acceptable solution should satisfy 
the criterion $\chi^2_{\min}(\hat{b}) \approx N_{df}$, where $\chi^2_{\min}$ is 
the value of the $\chi^2$ function at the minimum of \eq{F}. For the cutoff 
regularisation discussed later $N_{\rm df}=n_a-T$ is exact.

Regarding the posterior response matrix $P$, the trace $T$ is a measure of 
the residual smearing affecting $\hat{b}$. Noting that the elements $P_{kl}$ 
of a response matrix can be interpreted as probabilities that an event bin 
$l$ is observed in $k$, and since $T$ is a sum over the diagonal elements, 
the ratio $T/n_b$ is the  average fraction of events in which the unfolded 
value is in the same bin as the true value, and $1-T/n_b$ quantifies the 
amount of bin-to-bin migration in the unfolded distribution. Assuming a 
gaussian resolution function this can be translated into an average value 
for the posterior resolution. If the bin-width is equal to one standard 
deviation, then about 38.3\% of the measurements are not affected by 
bin-to-bin migration, and one would have $T \approx 0.4\,n_b$. If in a 
given case $T$ is significantly smaller, then one should consider 
rebinning the result.
 
The above considerations show that the unfolding result in general is 
still a distorted version of the truth. For the fit of a parametric
model to the unfolded result therefore the same caveats apply as for 
a fit to the original data. For fitting the original data, the 
model has to be forward folded with the response matrix $R$,
when fitting the unfolding result the model needs to be forward
folded with the posterior response matrix $P$.

We conclude this section by giving the explicit expressions for the 
posterior response matrix in the Tikhonov-regularisation scheme \eq{F}. 
The explicit form of $M$ follows from the condition $\partial F/\partial b_i=0$ 
that determines $\hat{b}$. Starting point is the total differential
\beq{}
0 = d\left(\frac{\partial F}{\partial b_i}\right)
  = \sum_{k=1}^{n_b} \frac{\partial^2 F}{\partial b_i\partial b_k} d\hat{b}_k
  + \sum_{l=1}^{n_a} \frac{\partial^2 F}{\partial b_i\partial a_l} d\hat{a}_l \;,
\eeq
with derivatives taken at the measured values $\hat{a}$ and the estimates 
$\hat{b}$. Introducing the second-derivative matrices $G$ and $H$ and 
switching to matrix notation leads to
\beq{}
   0 = H\,d\hat{b} +  G\,d\hat{a}
   \Text{with}
   H_{ik} = \frac{\partial^2 F}{\partial b_i\partial b_k}
   \Text{and}
   G_{il} = \frac{\partial^2 F}{\partial b_i\partial a_l} \;,
\eeq
and the matrix $M$ with elements $M_{kl} = d\hat{b}_k/d\hat{a}_l$ becomes 
\beq{}   
   M = -H^{-1}  G \;. 
\eeq
With $F= \chi^2 + w S$ as defined in \eq{F} the explicit expressions for 
the second-derivative matrices are
\beq{}
    H = 2\,R^T\,C^{-1}\, R + w\, S''
    \Text{and}
    G = -2\, R^T \,C^{-1} \;,
\eeq
where $S''$ is the Hessian of the smoothing term, so one finds
\beq{}
   M = \left(R^T C^{-1} R + \frac{w}{2} S''\right)^{-1} R^T C^{-1} 
\eeq
and the posterior response matrix $P=M\,R$ becomes
\beq{P}
   P = \left( I + \frac{w}{2} S''\right)^{-1} I 
  \Text{with}
   I = R^T\,C^{-1}\,R \;. 
\eeq
For constrained or over-constrained problems with $n_a\geq n_b$ the 
inverse of the matrix $I$ exists and is equal to the covariance matrix 
of the estimate $\hat{b}$ obtained from an unregularised fit with $w=0$. 
In this case \eq{P} can be rewritten as
\beq{PP}
   P = \left(I^{-1} \left( I + \frac{w}{2} S''\right)\right)^{-1}
     = \left( 1 + \frac{w}{2} I^{-1} S''\right)^{-1} \;.  
\eeq
Here ``1'' denotes the unit matrix. For $w=0$ the posterior response is perfect, 
for $w>0$ residual distortions occur. It is worth noting that while $I$ is a 
symmetric matrix, $P$ usually is not. As can be seen from \eq{PP}, in Tikhonov 
regularisation based on \eq{F} it is only symmetric if $S''$ is proportional 
to the unit matrix.

\section{The Fisher basis}
The naive basis in which to construct the unfolded distribution $\hat{b}$
is given by the individual components $b_j$, $j=1,\ldots,n_b$. The 
information content of a measurement $\hat{a}$ about the true distribution
$b$ can be quantified by the Fisher information matrix \cite{Fisher1922},
which according the Cram\'er-Rao bound is the inverse of the covariance matrix 
of the unbiased estimator $\hat{b}$ with minimum variance \cite{Cramer1946,Rao1945}. 
It thus is a measure for the attainable accuracy when the expectation value of the
result is equal to the truth. Here we consider gaussian or Poisson-distributed
measurements, where the Fisher information matrix is given by
\beq{}
           I = R^T C^{-1} R  \;.
\eeq

To analyse the problem further, it is advantageous to chose a basis for the 
unfolded distribution for which the expansion coefficients are statistically 
independent. This basis is given by the eigenvectors or the Fisher information 
matrix, which we will refer to as ``Fisher basis'' in the following. Since 
the matrix $I$ is by construction symmetric and positive definite, the 
eigenvectors form an orthonormal basis with
\beq{}
   V\,V^T = V^T\,V = 1\;.
\eeq
In the Fisher basis $V$ the true distribution $b$ is transformed to a 
vector $\alpha$
\beq{fisher}
     \alpha = V^T\,b 
\eeq
with diagonal information matrix
\beq{}
     V^T\, I\, V = E \;.
\eeq
Each diagonal element specifies the amount of information that the data 
contribute to the respective expansion coefficient. If the eigenvalue 
$E_{kk}$ is small, then the coefficient $\alpha_k$ is only weakly 
constrained by the data.

The Fisher basis, which is also used in SVD-based unfolding methods 
\cite{Hoecker1996}, has the advantage that in this representation the 
inverse problem becomes most transparent. The unregularised best-fit 
parameters $\hat{\alpha}^0$ can be read off from \eq{lsqfit} as
\beq{} 
     \hat{\alpha}^0 = V^T\,\hat{b}^0 
     \Text{with}
     \hat{b}^0 = (R^T C^{-1} R)^{-1}  R^T C^{-1} \hat{a} \;,
\eeq
and the covariance matrix becomes
\beq{ca0}
    C(\hat{\alpha}^0) = V^T C(\hat{b}^0) V = E^{-1} \;.
\eeq
The coefficients $\hat{\alpha}^0$ are uncorrelated with variances given by 
the inverse of the eigenvalues of the Fisher information matrix. Given 
$\hat{\alpha}^0$, regularisation can be implemented by multiplication with 
a damping matrix $D$, such that $\hat{\alpha}_k\approx \hat{\alpha}^0_k$ for 
well measured coefficients and $\hat{\alpha}_k \approx 0$ for coefficients 
that are not constrained by the measurements $\hat{a}$. This leads to the 
regularised coefficients  
\beq{Chatest}
   \hat{\alpha} = D \, \hat{\alpha}^0
   \Text{with covariance matrix}
   C(\hat{\alpha}) = D \, E^{-1} D^T  \;.
\eeq
The posterior response matrix for the unfolded distribution $\hat{b}$
\beq{}
    P = M \, R  \Text{with}  M_{ik} = \frac{d\hat{b}_i}{d\hat{a}_k}
    \Text{and} \hat{b} = V\,\hat{\alpha}
\eeq
is obtained as
\beq{}
   P = V \,D \, V^T \;.
\eeq
In the framework of Thikonov regularisation this result corresponds
to a smoothing term $wS(b) = b^T\,V\,E\,(D^{-1}-1)\,V^T\,b$.
Since $V$ is an orthogonal matrix one has $T = {\rm Tr}\,P ={\rm Tr}\,D$.
If the damping matric $D$ is symmetric, then also the posterior 
response matrix is symmetric, and if $D$ is diagonal, then in 
addition also the regularised coefficients $\hat{\alpha}$ are 
uncorrelated.

\section{Regularisation with a discrete-valued penalty function}
The above formalism usually employs a differentiable smoothing
function $S(b)$, which adds a penalty to the $\chi^2$-term when the 
solution $b$ deviates from the prior expectations coded into $S(b)$. 
A problem with the classical Thikonov regularisation \eq{F} is the need
to adjust the regularisation parameter $w$, where a given strategy often 
works but sometimes produces unsatisfactory results. To address this 
issue, we explore here the use of a discrete-valued penalty function, 
defined by 
\beq{DS}
   S(b) 
 = \sum_{i=2}^{n_b-1} 
   \left\{ \begin{array}{l} 
             0  \Text{if} (b_{i-1}-b_i)(b_{i+1}-b_i) \leq 0  \\
             6  \Text{if} (b_{i-1}-b_i)(b_{i+1}-b_i)  > 0  
           \end{array}
   \right.
\eeq
This penalty function scans all interior bins of $b$. If there is 
monotonic behaviour, i.e.~if $b_i$ is in-between its neighbouring values, 
there is no penalty. If $b_i$ is larger or smaller than both its 
neighbours, then a fixed penalty of 6 units is added. 

This ansatz exploits the fact that the $\chi^2$ function defines a natural 
metric for gauging the quality of a fit. The above penalty function allows 
the unfolding algorithm to remove a local extremum in the unfolded 
distribution if the modification in $\hat{b}$ increases the $\chi^2$ 
defined in \eq{chi2} by less than 6 units. The ``6'' is somewhat arbitrary, 
corresponding to the rationale that a spurious extremum is caused by a 
statistical fluctuation between 2 and 3 standard deviations, so that 
getting rid of it at the expense of increasing the $\chi^2$ between 4 
and 9 units is acceptable. 

It is worth emphasising that aside from the size of the penalty term, here
fixed to the value of 6, there is no regularisation parameter that needs to 
be adjusted. This is different for the regularisation parameter $w$ in \eq{F}, 
where the parameter $w$ balances the goodness-of-fit measure $\chi^2$ against 
the smoothing term $S$, which in typical application is a functional of the 
shape of the unfolded distribution. The parameter $w$ is needed as a 
conversion factor between $\chi^2$ and e.g.~curvature or entropy of $b$. 
In contrast to this, by \eq{DS} a penalty term is introduced that uses the 
same metric as the goodness-of-fit measure, which means that $\chi^2$ and 
$S$ are of the same nature. This removes the need for $w$. The size of the 
penalty for non-monotonic behaviour is like a kind of look-elsewhere effect 
and should slowly grow with the number of bins that are considered in order 
to account for the fact that more bins make it more likely to obtain large 
fluctuations. Since only a weak dependence is expected, in the following 
only a single value for the penalty is considered.

The main problem is to find the minimum of \eq{F} with a discontinuous 
regulariser \eq{DS}. Since any gradient-based method will fail, one 
possibility is a phase-space scan using an MCMC-technique, which, however, 
becomes computationally very expensive. In the following we will therefore 
explore the use of a discrete-valued penalty function by implementing a 
less ambitious approach, which may not find the global optimum but is 
expected to return a solution close to it. 
 
The starting point is to express the unfolding problem in the orthonormal
Fisher basis. The basis vectors are ordered such that the corresponding
eigenvalues, i.e. the inverse of the variances of the respective expansion 
coefficients, are given in decreasing order. The low-order eigenvectors 
are well measured. For response matrices describing a detector with finite 
resolution, those eigenvectors show little variations, which for a set of 
orthonormal vectors is equivalent to fewer numbers of zero crossings and 
thus to a lower numbers of local extrema. The strategy to construct 
a single estimate $\hat{b} = V \hat{\alpha}$ of
the unfolded distribution is as follows:

\begin{itemize}
\item Assume that all coefficients are in the range $\hat{\alpha}_k\in[-N,N]$, 
      where $N$ is the number of entries in the measurements $\hat{a}$ of the 
      observed distribution. This is a soft regularisation step. 
      It is found that deviating from the natural scale $N$ by a 
      factor of 2 in either direction has negligible impact on the result,
      too small values will heavily bias the result, too large values 
      lead to numerical instabilities.     

\item Initialise all coefficients to $\hat{\alpha}_k=0$.

\item Minimise in turn the cost function $F(\hat{\alpha})$ for all parameters.
      First find the parameter $\hat{\alpha}_k$ for $k=1$ that minimises 
      $F(\hat{\alpha})$ when all higher order parameters are zero. 
      Then fix $\hat{\alpha}_{k}$ to the value just found 
      and vary $\hat{\alpha}_{k+1}$ such that 
      $F(\hat{\alpha})$ is minimal. All higher order parameters are still zero. 
      Repeat the procedure until all parameters are determined.

\item For the minimisation a simple 1-dimensional iterative grid search is used. 
      In each step
      the cost function is evaluated at 25 equidistant points in the search 
      interval, then the location of the minimum is taken and the interval size
      reduced by a factor of 5, centered around the current minimum. These 
      steps are iterated until the interval size drops below 0.01.

\item The finite statistical precision of the measurements $\hat{a}$ is 
      accounted for by the bootstrap method \cite{Efron1986}. It is 
      realised by Poisson fluctuations of the bin contents $\hat{a}$. 
      Each fluctuation then is unfolded and the individual estimates 
      averaged to obtain the nominal unfolding result. The scatter of 
      the individual results determines the covariance matrix of the 
      unfolded distribution. 
\end{itemize}

\begin{figure}[tb]
\centering
\includegraphics[width=0.95\textwidth]{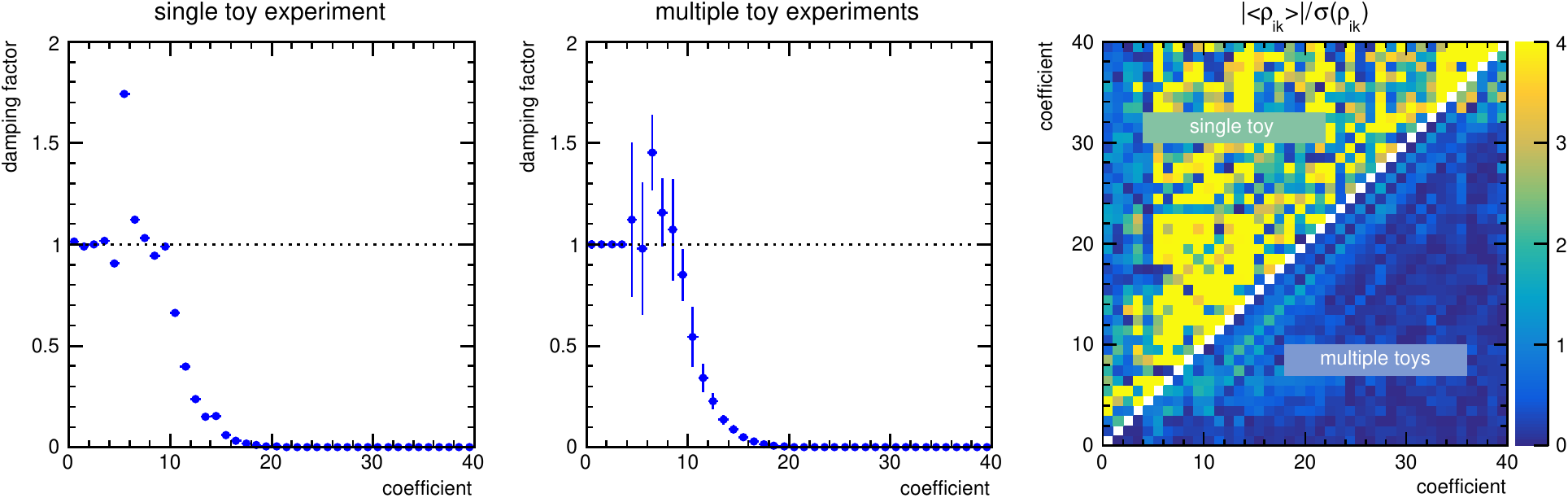}
\caption{Damping factors inferred from the diagonal elements of the 
         covariance matrix of the regularised expansion coefficients.
         Here a single covariance matrix estimate is constructed from 
         10\,000 bootstrap samples. The plots show mean value and standard 
         deviation from 100 independent estimates, the left hand plot 
         from the data of a single toy experiment, the middle one for 
         100 independent toy experiments. The right hand plot displays 
         the significance of the correlation coefficients between the
         damping factors, defined as the absolute value of the average 
         divided by the RMS error. The upper left triangle shows this 
         information for the single toy experiment, the lower right 
         triangle for the 100 independent toy data sets. The diagonal 
         is set to zero. While significant correlations are seen for a 
         single toy experiment, the average over many independent
         toy data sets appears largely uncorrelated.}
\label{fig:damping}
\end{figure}

The bootstrap approach is chosen both for its conceptual simplicity and 
since conventional error propagation based on derivatives of the result 
with respect to the inputs is not applicable because of the discontinuities 
in the cost function. The numerical results shown in section \ref{sect:toy} 
are based on 1000 bootstrap samples. Increasing the number to 10\,000 
entails no visible changes.

It remains to construct an estimate of the posterior response matrix,
which in the Tikhonov regularisation scheme is related to the damping  
matrix $D$ that is applied to the unregularised coefficients, \eq{Chatest}.
How to define a damping matrix for a discontinuous penalty function 
\eq{DS} is not at all obvious, but an effective damping matrix can be
constructed by exploiting the relation between the damping matrix and 
the covariance matrix of the unfolding result. 

When the effect of the regularisation is described by a damping matrix 
$D$ that is applied to the vector of the unregularised coefficients in 
the Fisher basis, the covariance matrix of the unfolded distribution is 
\beq{Cest}
    C(\hat{b}) = V\,D\,E^{-1}\,D^T\,V^T  \;,
\eeq
which relates a matrix $C(\hat{b})$ obtained from bootstrap variations to 
the damping matrix $D$. Here $V$ and $E$ are known. Since the expansion 
coefficients $\hat{\alpha}_k$ are statistically independent, $D$ can be 
expected to be dominated by the diagonal elements, which in turn can be 
estimated from the diagonal elements of an auxiliary matrix 
$X = V^T C(\hat{b}) V$ by
\beq{damp}
    D_{ii} = \left\{
    \begin{array}{l}
    \sqrt{E_{ii} X_{ii}} 
    \Text{if} 
    \sqrt{E_{ii} X_{ii}} <1 \quad\forall\quad k\geq i \\
     1 \Text{otherwise}
    \end{array}
    \right.
    \;.
\eeq
This condition implements the assumption of a diagonal damping matrix,
and avoids unphysical behaviour due to statistical fluctuations or
limitations of the diagonal approximation by taking the low order 
damping coefficients as unity up to the point where the estimate 
based on the diagonal elements of $X$ drops below unity. 

Numerical estimates of the damping factors in the toy model $f_1(y)$ 
presented in section \ref{sect:toy} are shown in fig.~\ref{fig:damping}. 
Here 10\,000 bootstrap samples are generated for a single estimate
of the covariance matrix $C(\hat{b})$, from which the damping factors are 
determined according to \eq{damp}. Uncertainties of those estimates are 
obtained from the RMS scatter of 100 independent such estimates. Doing 
these 100 estimates on a single toy experiment determines the statistical 
precision of a single bootstrap estimate. Generating a new toy experiment 
for each bootstrap estimate determines the actual uncertainty of the 
damping factors. Also shown in fig.~\ref{fig:damping} are the
significances of the correlations between the damping factors.
As expected, one finds significant correlations when keeping the 
pseudo-data sample fixed, while no significant correlations are observed
when varying those data. This corroborates the assumption of a diagonal 
damping matrix. 

The study also shows that within uncertainties the low-order damping 
factors are consistent with unity, even if the estimates for a single
sample may be significantly off. The sensitivity to fluctuations in 
the data is lower for the damping factor of the higher order 
coefficients, i.e. those can be reliably estimated also from a single
given data sample. \Eq{damp} appears to be a viable method to
obtain an estimate for the damping matrix $D$ and the posterior 
response matrix $P$. 

\begin{figure}[t]
\centering
\includegraphics[width=0.95\textwidth]{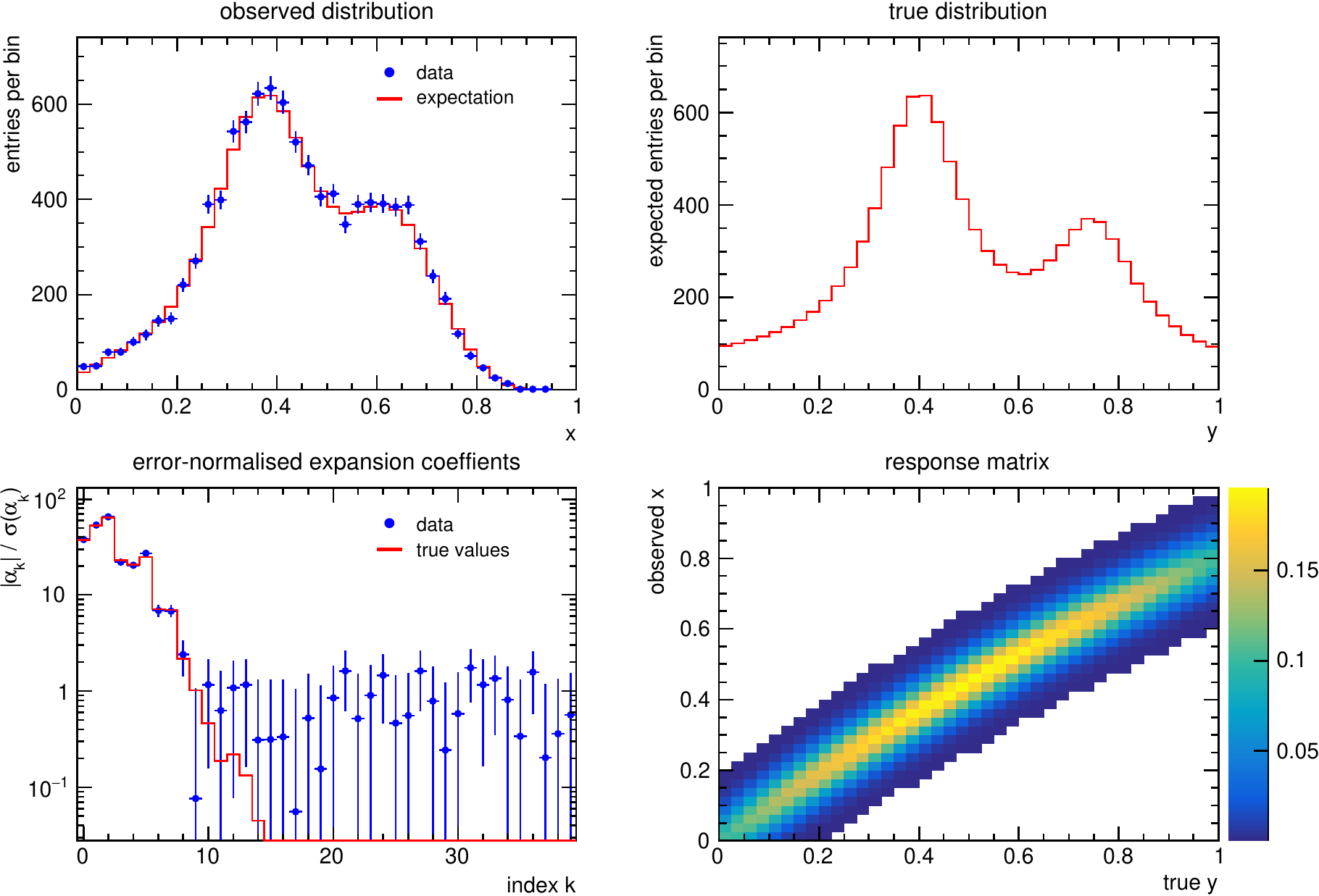}
\caption{\small Illustration of a discrete unfolding problem. The right 
          hand column shows the assumed true distribution and the 
          response matrix. Efficiency loss, bias and smearing are clearly 
          visible. The points with error bars in the top left plot are 
          pseudo data for an expected number of 10\,000 entries, the 
          histogram shows the expected bin contents. One sees how the  
          actual measurements fluctuate around their expectation values. 
          The bottom left plot illustrates how well the expansion coefficients 
          $\alpha_k$ in the Fisher basis can be measured. Shown are the 
          ratios $|\alpha_k|/\sigma(\alpha_k)$, i.e.~the significances of 
          the expansion coefficients, where histogram and points correspond 
          to the respective objects in the plot above. Beyond $k=O(10)$ no 
          statistically significant measurement of the $\alpha_k$ is possible.}
\label{fig:f1_data} 
\end{figure}

\section{Numerical studies}
\label{sect:toy}
The ansatz of a discrete-valued penalty function is tested in toy model
studies, where different generic true distributions are distorted by a 
response function that models non-uniform efficiency losses, biased 
measurements and gaussian resolution effects. In order to establish a 
baseline against which to gauge the performance of the new method, we 
will first discuss a simple cutoff regularisation and then switch to 
discrete-valued penalties. In all cases the true distribution 
is defined on the range $y\in[0,1]$, measurements are considered for 
$x\in[0,1]$. The response function is given by 
\beq{}
   R(x,y) 
 = \frac{1}{\sqrt{2\pi}\sigma} \bigg(\frac{1}{2}+2y(1-y)\bigg)
   \exp\left( -\frac{(x-  y - y^2/5)^2}{2\sigma^2}\right)  \;,
\eeq 
and the true PDFs considered are proportional to
\beq{toys}
    f_1(y) &= \frac{10}{100 +(10x-2)^2}
            + \frac{1}{1 +(10x-4)^2}
            + \frac{2}{4 +(20x-15)^2}  \\[4mm]
    f_2(y) &= 2 \exp\big(-200(x-0.35)^2\big)
              + \exp\big(-200(x-0.65)^2\big)\\[3mm]
    f_3(y) &= \left\{ \begin{array}{l}
                              1 \Text{for} y\in    [0.25,0.75] \\
                              0 \Text{for} y\notin [0.25,0.75] 
                             \end{array}\right. \\[1mm]
    f_4(y) &= \left\{ \begin{array}{ll}
                              \exp(1-5y) &\mbox{for}\quad y\geq 0.2 \\
                              0          &\mbox{for}\quad y<0.2
                             \end{array}\right. \;.
\eeq
The response function $R(x,y)$ and $f_1(y)$ correspond to the prototype
unfolding problem introduced in reference \cite{Blobel1984}. The response 
function describes a parabolic efficiency function with 100\% at $y=1/2$ 
that drops to 50\% at $y=0$ and $y=1$, a non-linear bias that grows 
from zero at $y=0$ to $0.2$ at $y=1$, and a gaussian smearing.
The resolution parameter is $\sigma=0.05$. The function $f_1(y)$ 
represents two Breit-Wigner peaks on top of wider background density 
that is also parametrised by a Breit-Wigner function. The alternative 
distributions realise a simple two-mode density of two gaussian peaks, 
a box-function and an exponentially falling spectrum.

\begin{figure}[t]
\centering
\includegraphics[width=0.95\textwidth]{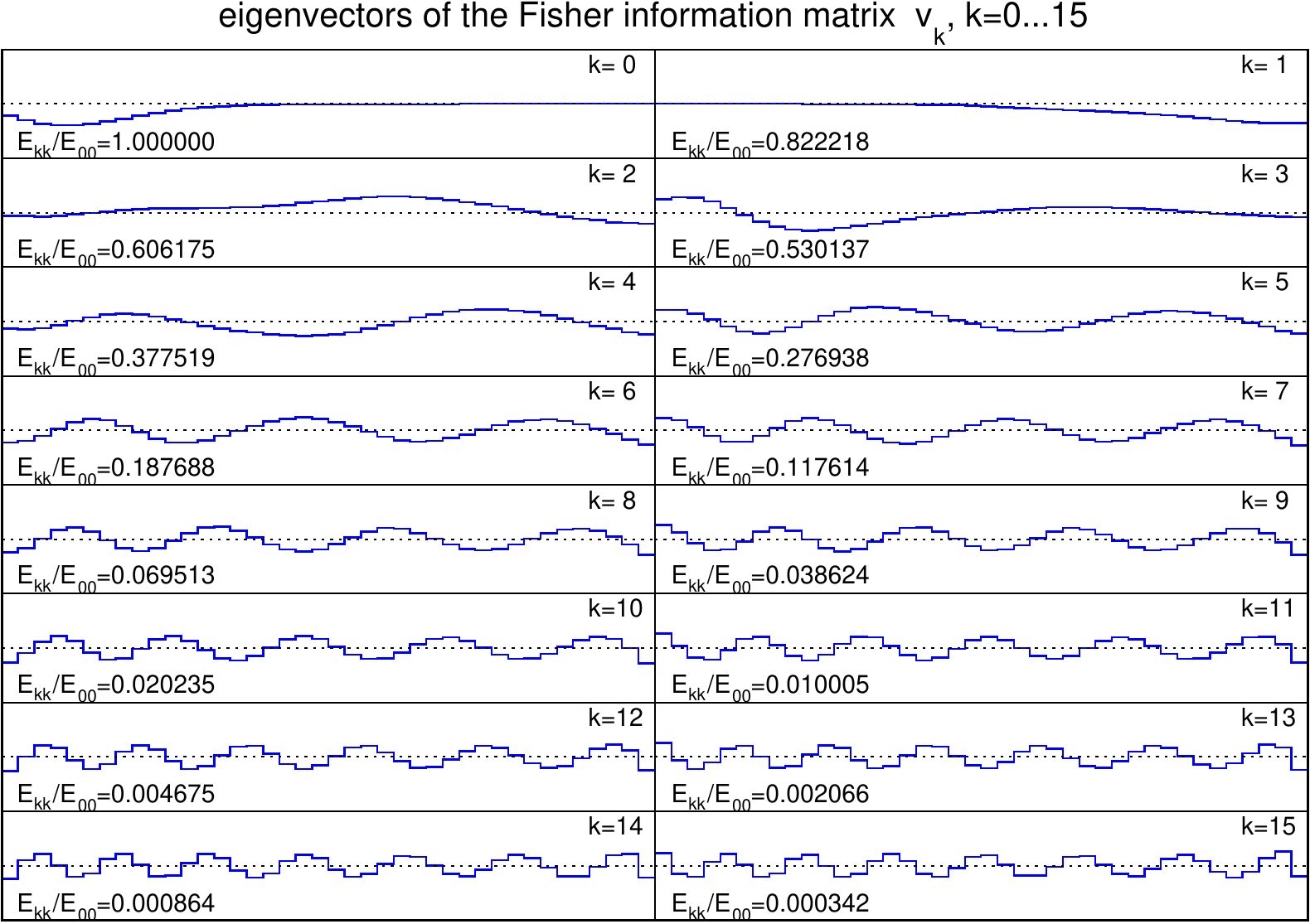}
\caption{\small Fisher basis for the unfolding problem shown in 
         fig.~\ref{fig:f1_data}. The ratios $E_{kk}/E_{00}$ quantify
         by how much response functions does suppress the higher
         order eigenvectors compared to the leading order one.} 
 \label{fig:f1_fisher} 
\end{figure}

Since we are considering a discrete unfolding problem, the true distributions
are represented by $n_b$ equal-size bins over $y\in [0,1]$, which are mapped 
by the response matrix to $n_a$ observed bins over $x\in [0,1]$. The response 
matrix $R$, true distribution $b$ and the expectation values $a$ of the 
measurements are calculated according to \eq{disc}. The distributions
are normalised such that sum over all bins of the observed distribution
satisfies $\sum_{i=1}^{n_a} a_i=N$, where $N$ is the expected statistics 
for a given experiment. An actual measurement finally is generated by drawing 
for each bin a Poisson-distributed random variate around the respective 
expectation value.

Figure~\ref{fig:f1_data} illustrates the setup for the toy example $f_1(y)$. 
Both the true and the observed density are discretised by 40 bins. The figure 
shows how the response matrix applied to the true distribution determines the 
expectation values for the bins of the observed distribution, and how the 
measurements of a toy experiment with an expected number of 10\,000 entries 
scatter around their expectation values. Also shown are the significances, 
i.e.~absolute value divided by uncertainty, with which the coefficients of the 
expansion of the true distribution into the Fisher basis can be determined 
from the measurements. With an expected number of 10\,000 entries only the 
leading $O(10)$ coefficients are well determined. The higher order 
contributions are not accessible. Figure \ref{fig:f1_fisher} displays 
the leading 16 basis vectors of the Fisher basis. 

\begin{figure}[t]
\centering
\includegraphics[width=0.95\textwidth]{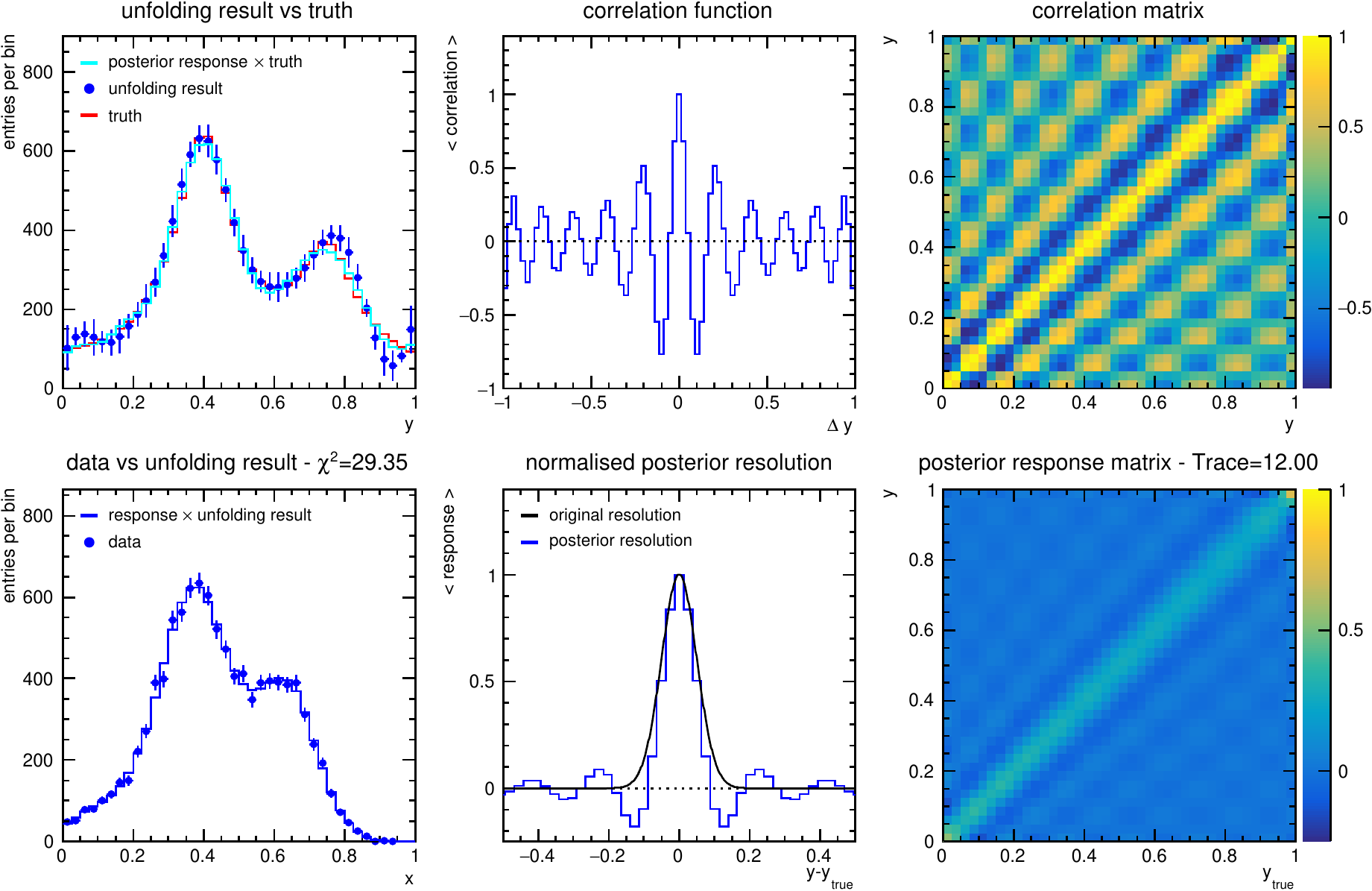}
\caption{\small Unfolding by using only the leading 12 coefficients of the 
          Fisher basis to construct an estimate of the true distribution.
          The top left plot is the unfolded distribution compared to the 
          truth and to the truth convolved with the posterior response.
          The posterior response matrix is shown in the bottom right plot.
          The top right plot displays the correlation matrix. The middle column 
          displays the correlation function \ref{eq:cdelta} (top) and posterior 
          resolution \ref{eq:pdelta} (bottom). The posterior resolution is 
          normalised to unity at $y-y_{\rm true}=0$. The bottom left, finally, 
          shows a comparison between the data and the unfolding result convolved 
          with the original response matrix show in fig.~\ref{fig:f1_data}. 
          The $\chi^2$ value tests the consistency between data and unfolding 
          result.} 
\label{fig:f1_leading12} 
\end{figure}

The simplest way to construct an estimate of the true distribution is by
cutoff-regularisation, namely to keep only the well measured leading coefficients 
and to synthesise the corresponding density by simply adding the corresponding 
basis vectors. Figure~\ref{fig:f1_leading12} shows the result when using the 
12 leading terms. Although a sharp cut on the number of terms in an expansion 
into orthogonal functions has the tendency to induce unwanted oscillations, 
this appears not to happen in this particular case. Evidently, the manual 
choice of the coefficients that are used to construct an estimate for 
the unfolded distribution introduces a subjective element into the procedure,
and there is a certain freedom to chose a solution as long as the result is 
statistically compatible with the uncorrected data. One possible criterion is 
the $\chi^2$ calculated according to \eq{chi2} under the assumption that the 
unfolding result $\hat{b}$ is the true distribution. The number of degrees of 
freedom for this $\chi^2$-value is $N_{\rm df}=n_a-T$, the number of bins $n_a$ 
used to represent the measured distribution minus the number of coefficients 
that contribute to the estimate $\hat{b}$ as given by $T={\rm Tr}Q ={\rm Tr}P$, 
with $Q$ the regularisation and $P$ the posterior response matrix. In addition 
it is suggested to provide also $P$, which quantifies to which extent the 
unfolding result $\hat{b}$ is still distorted as a consequence of the fact 
that any regularisation precludes a full correction of the detector response, 
and the matrix $\rho$ of the correlation coefficients between the bins of the 
unfolded distribution.

\begin{figure}[tb]
\centering
\includegraphics[width=0.95\textwidth]{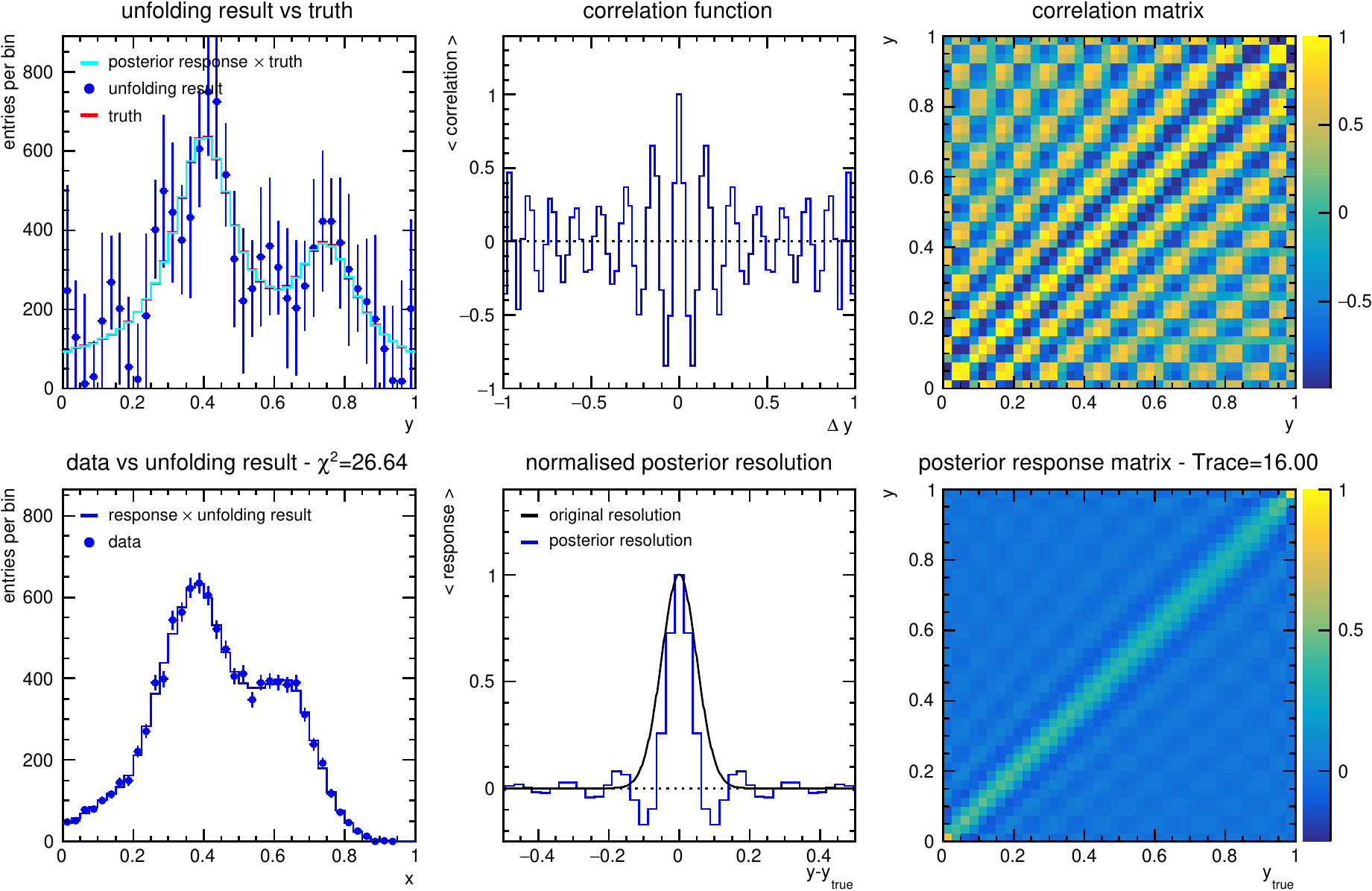}
\caption{\small Same as fig.~\ref{fig:f1_leading12} but using the 
          leading 16 coefficients of the Fisher basis to construct an
          estimate of the true distribution.}
\label{fig:f1_leading16} 
\end{figure}

The information content of the correlation matrix and the posterior 
response matrix can be visualised by the correlation function $c(\Delta)$ 
and the posterior resolution $p(\Delta)$, defined as the average correlation 
coefficient and the average posterior response as a function of the distance
$\Delta$ between two bins, 
\beq{cdelta}
     c(\Delta) 
  = \frac{1}{n_b-|\Delta|} \sum_{i,j=1}^{n_b} \rho_{ij} \delta_{i-j,\Delta}
\eeq
and
\beq{pdelta}
    p(\Delta) 
  = \frac{n_b}{T(n_b-|\Delta|)} \sum_{i,j=1}^{n_b} P_{ij} \delta_{i-j,\Delta} \;.
\eeq
For $p(\Delta)$ the normalisation $n_b/T$ in ensures $p(0)=1$. 

The top right plot of fig.~\ref{fig:f1_leading12} shows the correlation matrix, 
the top middle plot the correlation function. One observes strong and only 
slowly decaying anti-correlations between neighbouring regions, which implies 
that fluctuations in the data do not just affect a single bin of the unfolding 
result but can result in an oscillatory behaviour of the whole solution.

\begin{figure}[tb]
\centering
\includegraphics[width=0.8\textwidth]{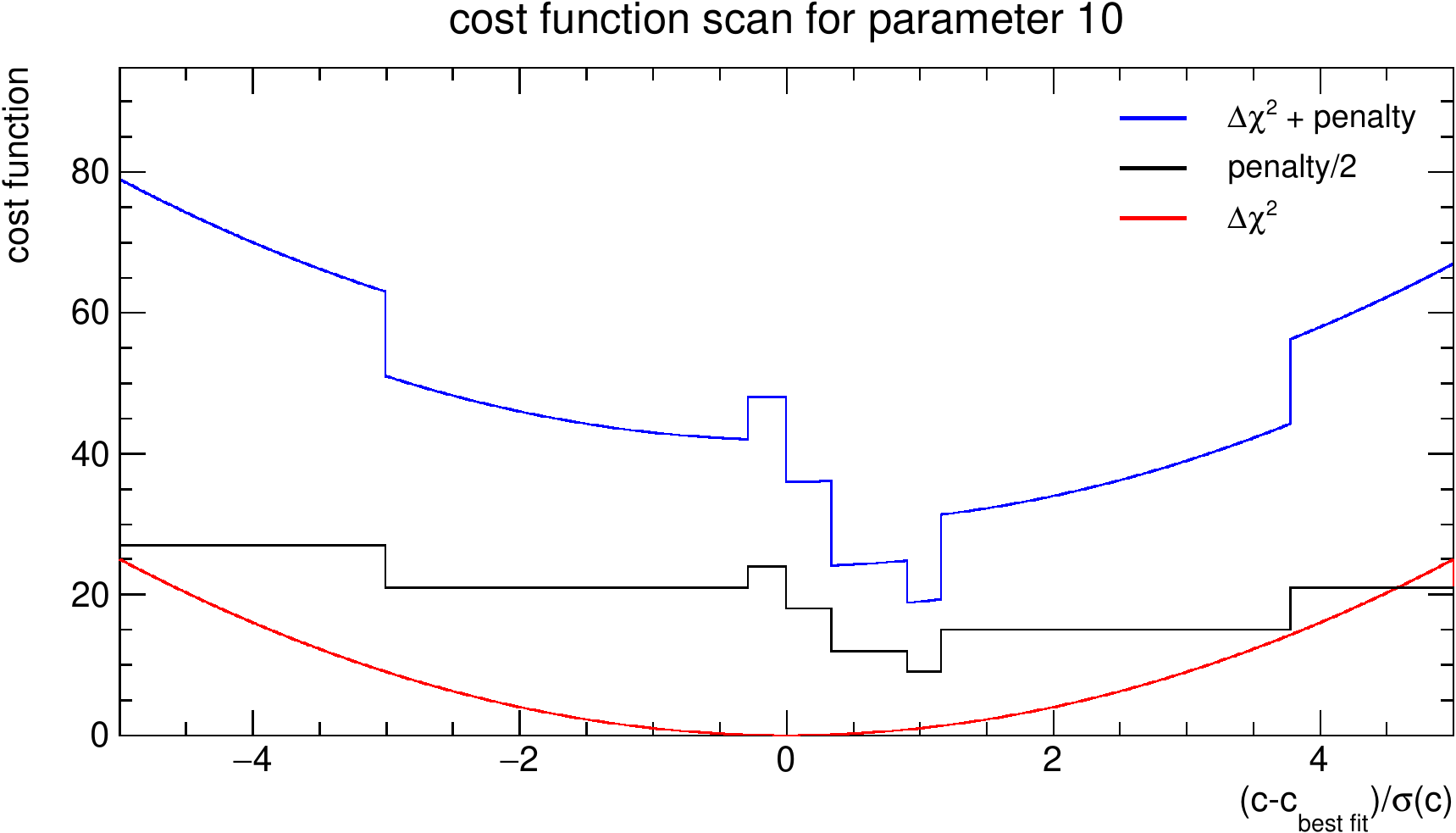}
\caption{\small Example for the cost function that has to be minimised 
         in order to determine one particular expansion parameter.
         Shown are $\chi^2$ contributions, the penalty function and 
         their sum.}       
\label{fig:f1_scan11} 
\end{figure}

The posterior response matrix in the lower right plot of 
fig.~\ref{fig:f1_leading12} shows that efficiency losses and biases of the 
measured $x$ compared to the true $y$ are corrected, but that the unfolding 
result is still smeared compared to the truth. It is symmetric about and 
homogenous along the diagonal. Looking at the posterior resolution, shown 
as a function of $y-y_{\rm true}$ in the bottom middle of 
fig.~\ref{fig:f1_leading12}, one sees that in this example the unfolding 
procedure did only marginally improve on the initial resolution. If the true 
function is sufficiently smooth, then the residual smearing will entail 
negligible distortions and the unfolding result will look as if the 
posterior response matrix were a unit matrix. In other words, the estimate 
$\hat{b}$ of a sufficiently smooth true distribution $b$ will have negligible 
bias, even though it is still a smeared version of the truth. The actual bias 
of the unfolding result depends on the true distribution, and if the truth is 
unknown, so is the bias. The posterior response matrix, however, allows one 
to calculate the bias for any assumed true distribution. This is also 
illustrated in fig.~\ref{fig:f1_leading12}, where the unfolding result is 
compared to the truth convolved with the posterior response.

Figure~\ref{fig:f1_leading16} illustrates how hard it is to correct for 
finite resolution effects. Using 16 instead of the leading 12 terms improves  
visibly the posterior resolution, but clearly destabilises the unfolding
result. Improving the resolution is only possible with more data, which then
provide the necessary statistical precision to determine also higher order
coefficients. 

This conceptual problem affects all regularisation methods. If there is
too little regularisation, then the posterior resolution is good, but the 
unfolding result is unstable. For stronger regularisation the unfolding 
result will stabilise, but smearing effects with respect to the truth are 
larger. In the language of the Fisher basis, the difference between 
different regularisation schemes translates into different approaches for
adjusting expansion coefficients that are not well constrained by the available 
data. It is important to keep in mind that the quality of an unfolding result 
cannot be judged by only looking at the estimate of the unfolded distribution.
It is mandatory to verify that the estimate convolved with the response 
matrix is consistent with data, and to give the posterior response matrix, 
or, at the very least, its trace, which quantifies the average posterior 
smearing.    

\begin{figure}[tb]
\centering
\includegraphics[width=0.95\textwidth]{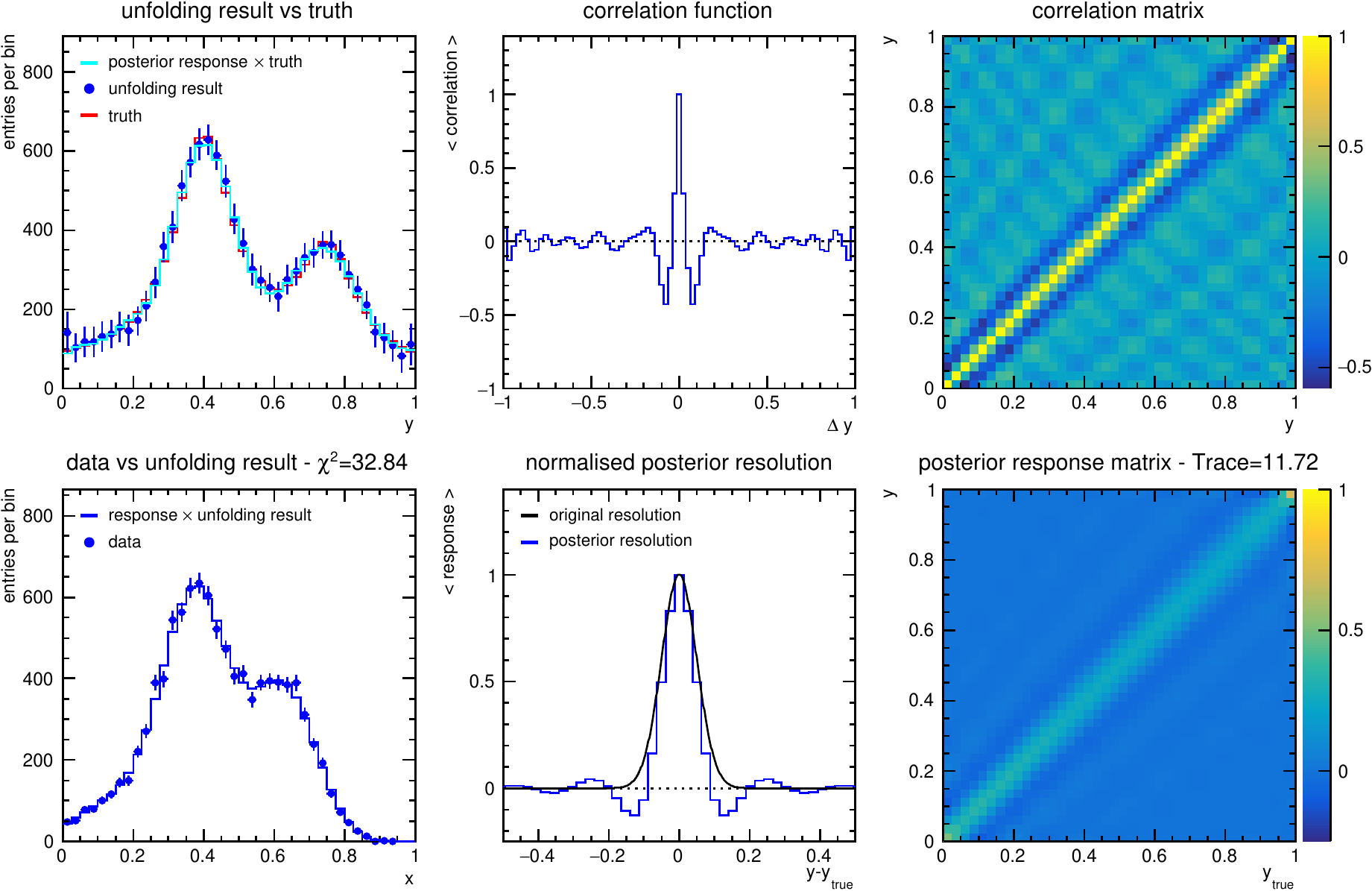}
\caption{\small Same as fig.~\ref{fig:f1_leading12} but using regularisation
          by a discrete-values penalty function to construct an estimate 
          of the true distribution.}
\label{fig:f1_dpr} 
\end{figure}

It is now interesting to see the performance of the regularisation by 
the discrete-valued penalty function. The algorithm used to test the concept 
as described before requires the minimisation of discontinuous functions in 
order to determine the expansion coefficients of the result in the Fisher 
basis. An example for such a function is shown in fig.~\ref{fig:f1_scan11}, 
which also gives a breakdown of the contributions to the total cost function.
The result of regularisation by the discrete-valued penalty function
\eq{DS} is shown in fig.~\ref{fig:f1_dpr}. Compared to unfolding by the 12
leading terms, a smoother result is obtained with improved posterior 
resolution and a posterior response function with less undershoot. 

The most dramatic change, however, is in the covariance matrix
of the result. The regularisation by the discrete-valued penalty function 
yields significantly reduced correlations, which intuitively can be 
understood by the fact that strong anti-correlations, which are typical
for unfolding problems, lead to local extrema in the result. Using a
regularisation that explicitly counters local extrema thus also counters
correlations. Another point worth mentioning is the behaviour of the
$\chi^2$ values for the consistency between unfolding result and data.
For a cutoff regularisation, where the leading order expansion coefficients
are taken at face value, the $\chi^2$ value is determined by setting the 
higher order coefficients to zero. For a discrete-valued regularisation 
also the higher order coefficients contribute to the result, which in principle 
should lead to reduction of the overall $\chi^2$ value. On the other hand,
now also the leading order coefficient can be varied in order to remove 
local extrema, which would entail an increase in the $\chi^2$ value. 
The net effect will depend on the case at hand. The crucial point, however, 
is that the final estimate for the unfolded distribution corresponds to a
statistically acceptable fit of the data.

\begin{figure}[p]
\centering
\includegraphics[width=0.95\textwidth]{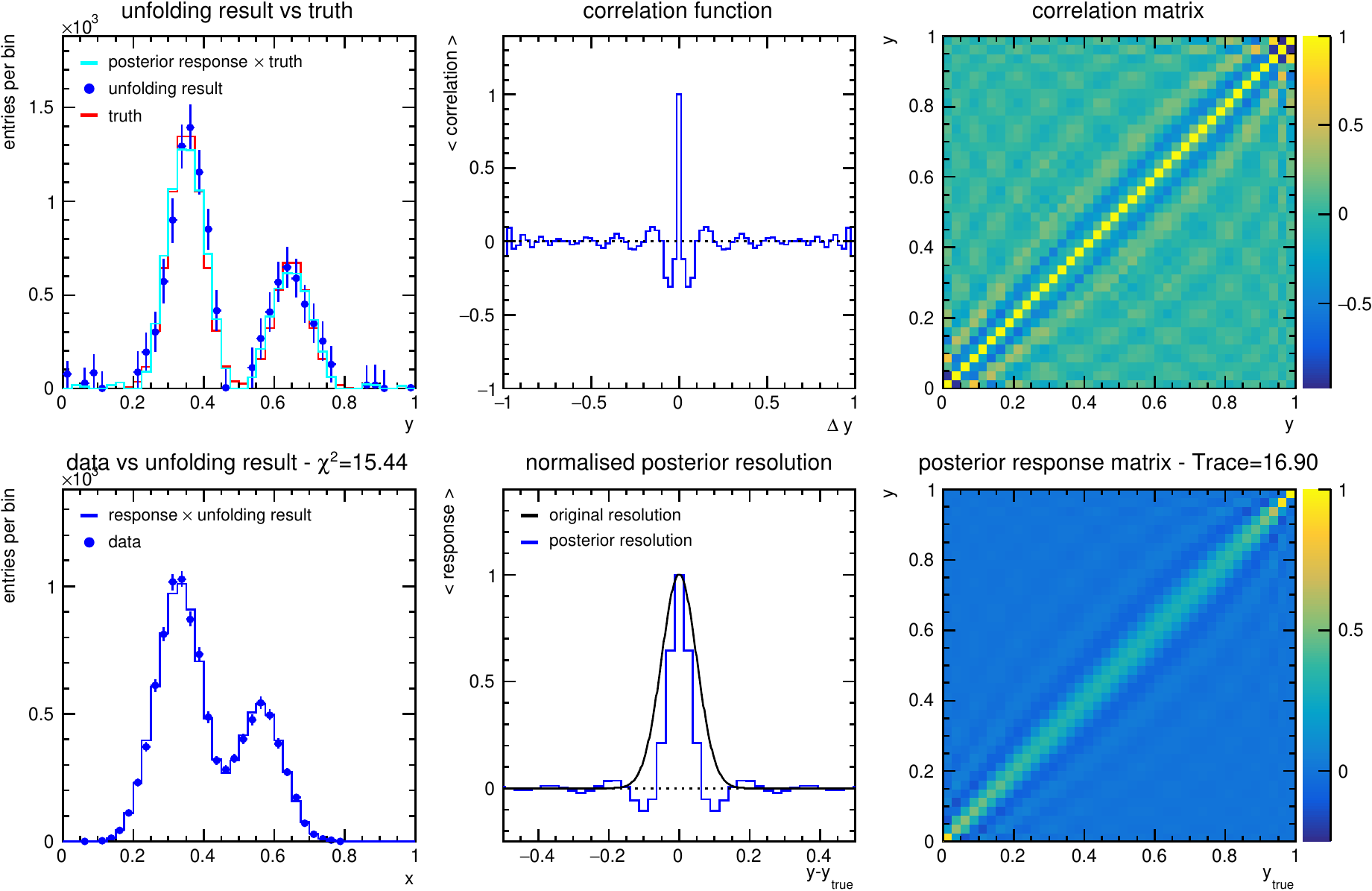}
\caption{\small Same as fig.~\ref{fig:f1_dpr} but for toy example $f_2(y)$.} 
\label{fig:f2_dpr} 
\vspace{10mm}
\includegraphics[width=0.95\textwidth]{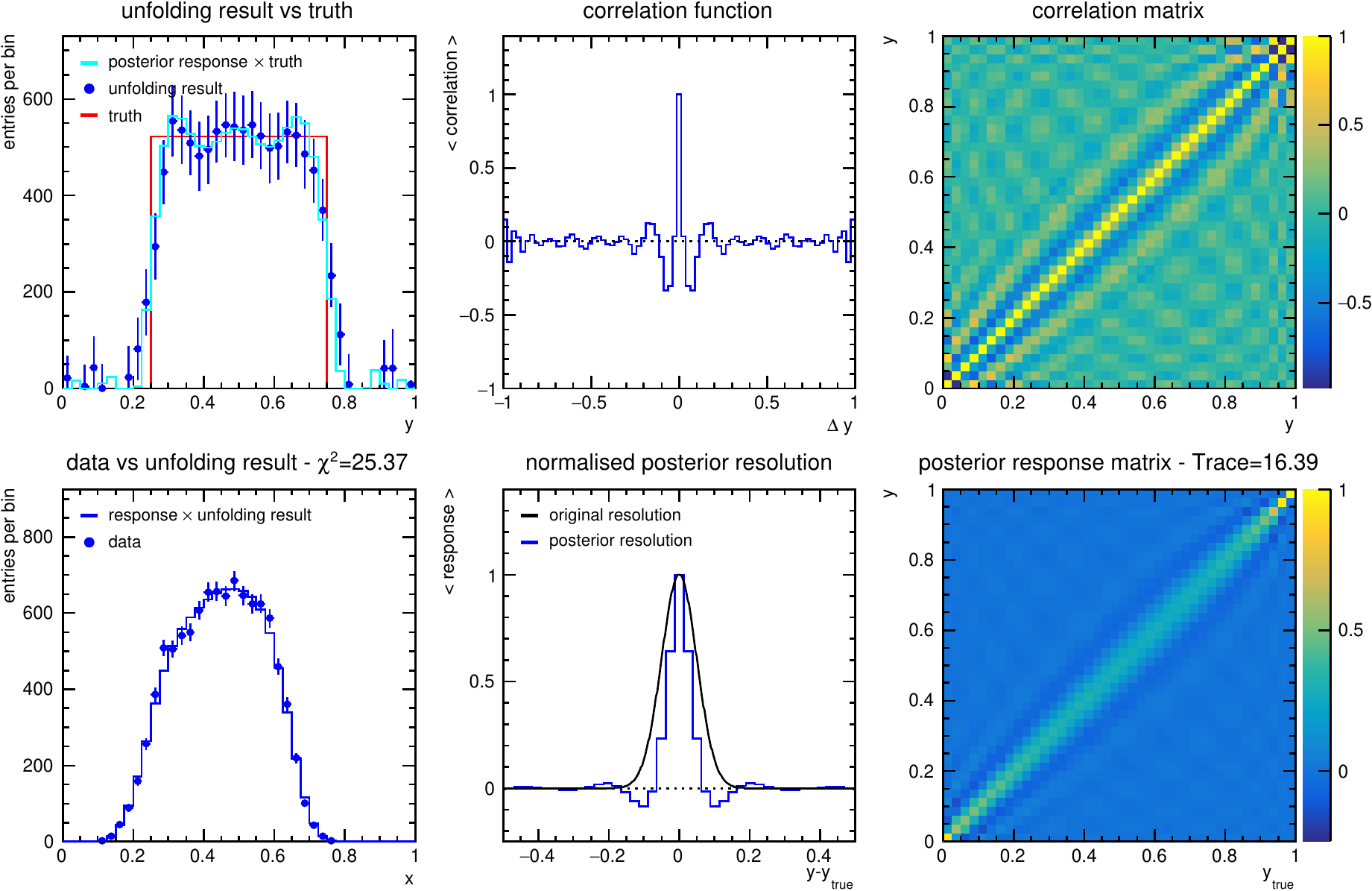}
\caption{\small Same as fig.~\ref{fig:f1_dpr} but for toy example $f_3(y)$.} 
\label{fig:f3_dpr} 
\end{figure}

\begin{figure}[t]
\centering
\includegraphics[width=0.95\textwidth]{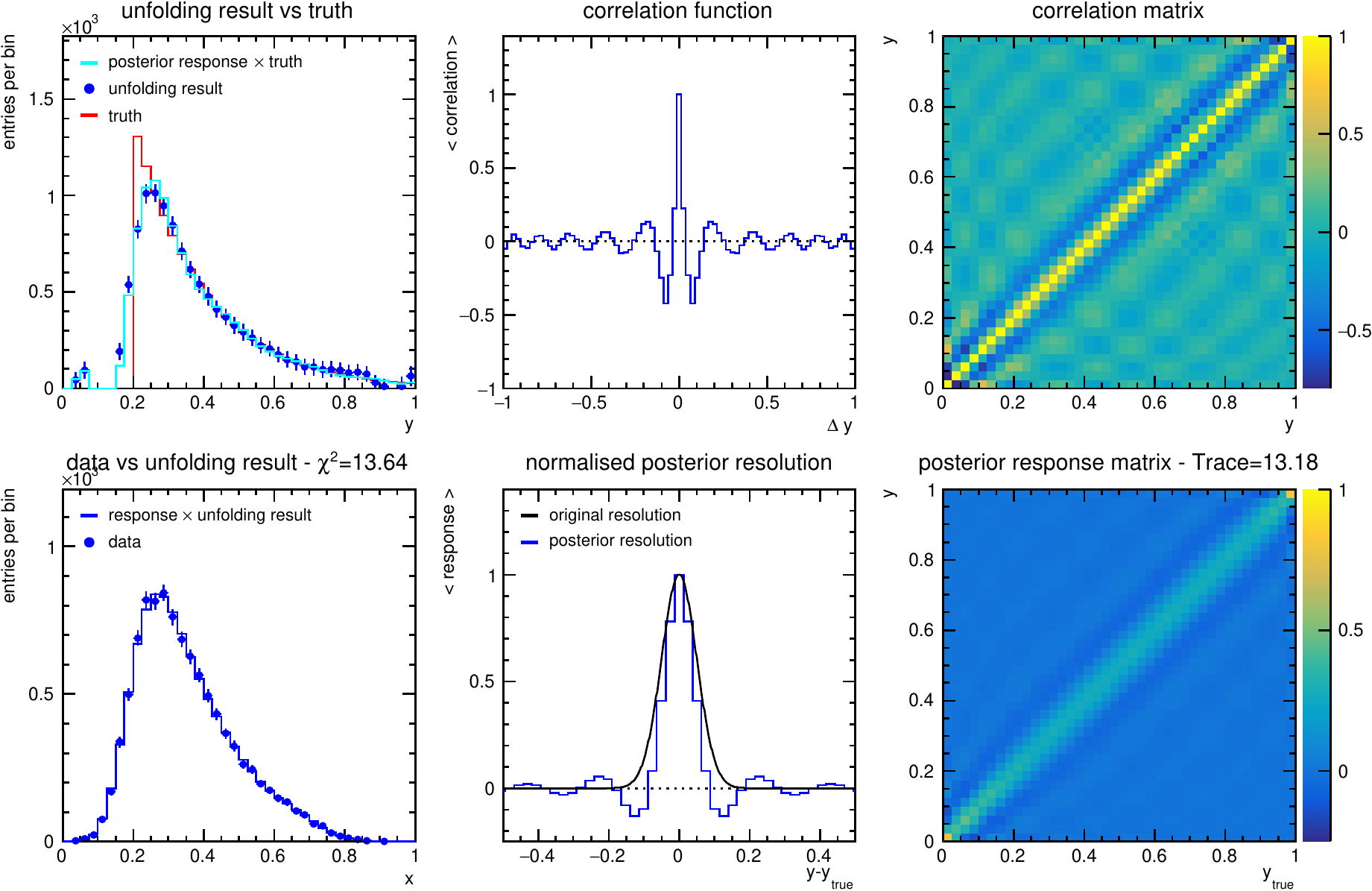}
\caption{\small Same as fig.~\ref{fig:f1_dpr} but for toy example $f_4(y)$.} 
\label{fig:f4_dpr} 
\end{figure}

\begin{figure}[t]
\centering
\includegraphics[width=0.95\textwidth]{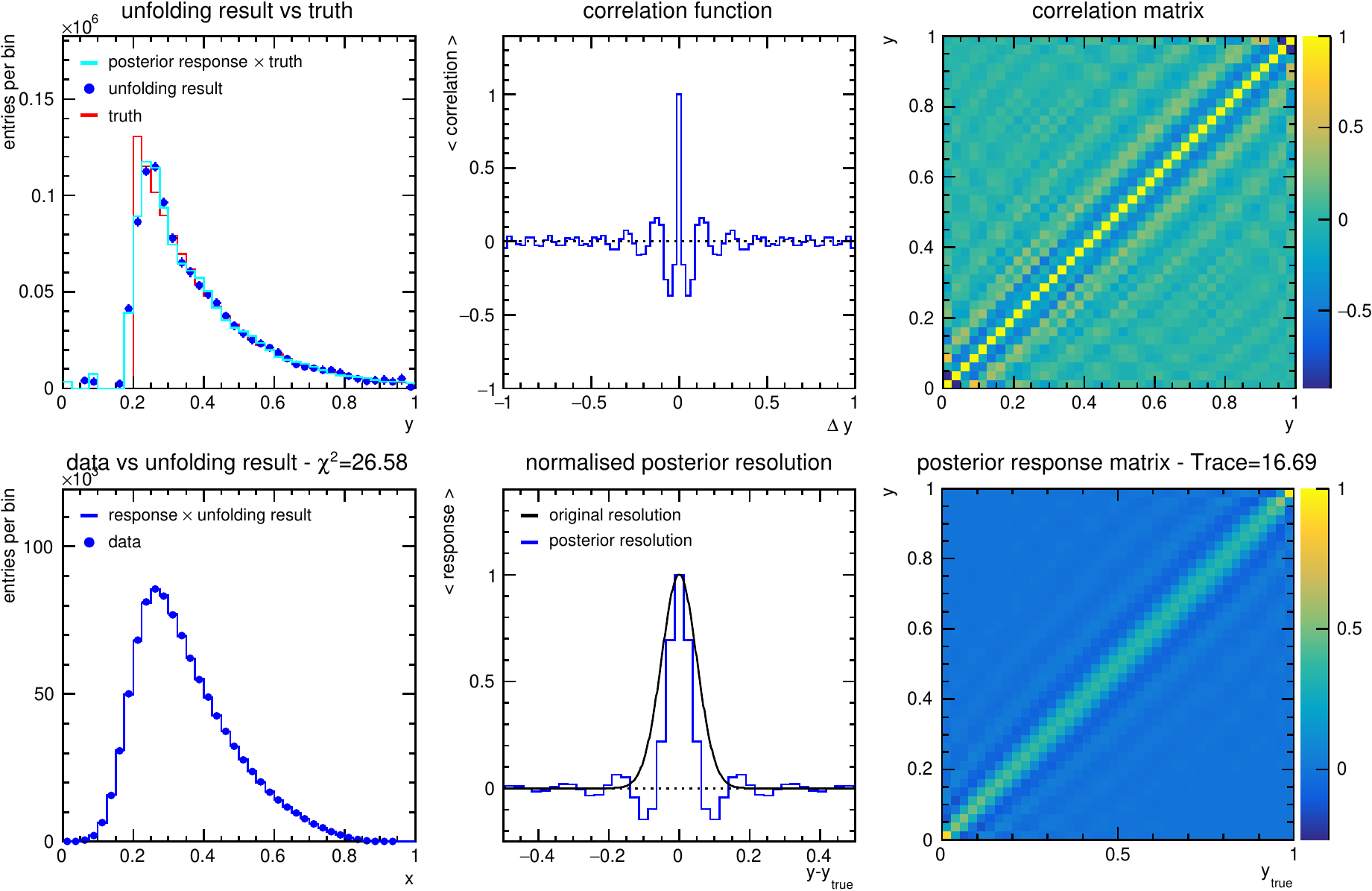}
\caption{\small Same as fig.~\ref{fig:f4_dpr} but with an expected number of 
         measurements of $10^6$ instead of $10^4$.} 
\label{fig:f4_dpr2} 
\end{figure}

It remains to test how regularisation by a discrete-valued penalty function
works for the other examples given in \eq{toys}. Those are presented in 
figs.~\ref{fig:f2_dpr}, \ref{fig:f3_dpr} and \ref{fig:f4_dpr}. In all
cases one gets quite satisfactory results without having to adjust a 
regularisation parameter. One also sees that the posterior resolution 
depends on the kind of problem one is solving, and that in all cases 
the truth convolved with the posterior response matrix agrees with 
the actual estimate of the true distribution within the calculated 
uncertainties. The most interesting test cases are shown in figs. 
\ref{fig:f3_dpr} and \ref{fig:f4_dpr}, where the true distribution exhibits 
discontinuities. The coefficients of the high order basis functions that 
would be needed to describe the edges usually cannot be determined from 
the available data with sufficient statistical precision. Sharp features 
are therefore most strongly affected by the regularisation, and it is 
reassuring to see that the estimates $\hat{b}$ nicely track the expectation 
when convolving the true distributions with the posterior response matrix.
Figure~\ref{fig:f4_dpr2} finally shows how the algorithm performs for 
the exponentially falling spectrum $f_4(y)$ when the statistics is 
increased from $10^4$ to $10^6$ expected events. Without any retuning the 
unfolding result has an improved posterior resolution, which reflects 
the ability to include higher order expansion coefficients.

\section{Summary}
We have discussed the discrete linear unfolding problem for the case that 
the response matrix and the covariance matrix of the measurements are known. 
The inverse problem to infer the true distribution from a finite statistics 
measurement is usually ill-posed, with the consequence that a solution 
always needs some kind of regularisation. A key element to quantify the 
effect of the regularisation is the posterior response matrix. While, as 
a consequence of the regularisation procedure, the unfolding result is still 
distorted with respect to the true distribution, it should be unbiased when 
compared to the true distribution convolved with the posterior response 
matrix. In numerical studies this can be checked, in real world applications 
one key criterion to test the validity of an unfolding result is to verify 
that the result when convolved with the original response matrix is 
statistically consistent with the measurements. Other criteria by which to 
judge an unfolding result are such as the stability of the result, properties 
of its error matrix or known physical constraints like positivity.

A special basis into which to expand the unfolded distribution is the 
Fisher basis, defined by the eigenvectors of the Fisher information
matrix, which quantifies the information content of the observed 
distribution about the bins of the true distribution. In the Fisher 
basis the unfolding problem is diagonal. If the regularisation acts 
independently on the individual eigenvectors, then the posterior
response matrix is symmetric.
 
The Fisher basis also provides a convenient starting point for implementing
a regularisation method, which uses a discrete-valued penalty function to 
suppress instabilities in the solution. We have presented such a method where 
the penalty is given in units of $\chi^2$. As a consequence there is no need 
for a case-by-case adjustment of a regularisation parameter, which in most 
commonly used methods is required to put goodness-of-fit and a suitably
chosen smoothness criterion on equal footing. 

For toy models a very satisfactory performance is found when using a 
discrete-valued penalty function. The results are characterised by how well 
they fit the data, and by their posterior response matrix. The posterior 
response allows one to estimate the conditional bias of the result, 
i.e.~the bias for an assumed truth, while the actual bias depends on the 
usually unkwown truth. Goodness-of-fit and posterior response are criteria 
that equally apply to all unfolding schemes. A quantitative comparison of 
results from different methods should be based on those and possibly on how 
the methods select subspaces from the full space of acceptable solutions. 
The toy studies also show clearly that while regularised unfolding can 
correct for biases and efficiency losses in the response function, only a 
partial correction for smearing effects is obtained. 

\section*{Acknowledgements}
It is a pleasure to thank Nikolay Gagunashvili from the University of Iceland
in Reykjavik for many inspiring and constructive discussions, and for the 
careful reading of the manuscript. Special thanks go to the anonymous
reviewer for the competent and constructive feedback during the review.

\end{document}